\documentclass[aps,prd,showpacs,amsmath,amssymb]{revtex4}

\usepackage{slashed}
\usepackage[dvips]{color}
\usepackage[dvips]{epsfig}
\usepackage{latexsym}
\usepackage{bm}
\usepackage{upgreek}
\usepackage{mathrsfs}
\usepackage{times}
\usepackage{amsthm}
\usepackage{amssymb}
\usepackage{epsfig}
\usepackage{graphicx}
\usepackage{amsmath}

\usepackage{color}

\definecolor{purple}{rgb}{0.8,0,0.6}

\begin{document}

\title{Strong dynamics behind the formation of the $125$ GeV Higgs boson }

\author{M.A.~Zubkov\footnote{on leave of absence from ITEP, B.Cheremushkinskaya 25, Moscow, 117259, Russia}}
\affiliation{University of Western Ontario,  London, ON, Canada N6A 5B7}

\begin{abstract}
We consider the scenario, in which the new strong dynamics is responsible for the formation of the $125$ GeV Higgs boson. The Higgs boson appears as composed of all known quarks and leptons of the Standard Model. The description of the mentioned strong dynamics is given using the $\zeta$ - regularization. It allows to construct the effective theory without ultraviolet divergences, in which the $1/N_c$ expansion works naturally. It is shown, that in the leading order of the $1/N_c$ expansion the mass of the composite $h$ - boson is given by $M_h = m_t/\sqrt{2} \approx 125$ GeV, where $m_t$ is the top - quark mass.
\end{abstract}



\maketitle


\newcommand{\br}{{\bf r}}
\newcommand{\bu}{{\bf \delta}}
\newcommand{\bk}{{\bf k}}
\newcommand{\bq}{{\bf q}}
\def\({\left(}
\def\){\right)}
\def\[{\left[}
\def\]{\right]}

\newcommand{\barray}{\begin{eqnarray}}
\newcommand{\earray}{\end{eqnarray}}
\newcommand{\nn}{\nonumber \\}
\newcommand{\nl}{& \nonumber \\ &}
\newcommand{\bnl}{\right .  \nonumber \\  \left .}
\newcommand{\dbnl}{\right .\right . & \nonumber \\ & \left .\left .}

\newcommand{\beq}{\begin{equation}}
\newcommand{\eeq}{\end{equation}}
\newcommand{\ba}{\begin{array}}
\newcommand{\ea}{\end{array}}
\newcommand{\bea}{\begin{eqnarray}}
\newcommand{\eea}{\end{eqnarray} }
\newcommand{\be}{\begin{eqnarray}}
\newcommand{\ee}{\end{eqnarray} }
\newcommand{\bal}{\begin{align}}
\newcommand{\eal}{\end{align}}
\newcommand{\bi}{\begin{itemize}}
\newcommand{\ei}{\end{itemize}}
\newcommand{\ben}{\begin{enumerate}}
\newcommand{\een}{\end{enumerate}}
\newcommand{\bc}{\begin{center}}
\newcommand{\ec}{\end{center}}
\newcommand{\bt}{\begin{table}}
\newcommand{\et}{\end{table}}
\newcommand{\btb}{\begin{tabular}}
\newcommand{\etb}{\end{tabular}}
\newcommand{\bvec}{\left ( \ba{c}}
\newcommand{\evec}{\ea \right )}

\newcommand\e{{e}}
\newcommand\eurA{\eur{A}}
\newcommand\scrA{\mathscr{A}}

\newcommand\eurB{\eur{B}}
\newcommand\scrB{\mathscr{B}}

\newcommand\eurV{\eur{V}}
\newcommand\scrV{\mathscr{V}}
\newcommand\scrW{\mathscr{W}}

\newcommand\eurD{\eur{D}}
\newcommand\eurJ{\eur{J}}
\newcommand\eurL{\eur{L}}
\newcommand\eurW{\eur{W}}

\newcommand\eubD{\eub{D}}
\newcommand\eubJ{\eub{J}}
\newcommand\eubL{\eub{L}}
\newcommand\eubW{\eub{W}}

\newcommand\bmupalpha{\bm\upalpha}
\newcommand\bmupbeta{\bm\upbeta}
\newcommand\bmuppsi{\bm\uppsi}
\newcommand\bmupphi{\bm\upphi}
\newcommand\bmuprho{\bm\uprho}
\newcommand\bmupxi{\bm\upxi}

\newcommand\calJ{\mathcal{J}}
\newcommand\calL{\mathcal{L}}

\newcommand{\notyet}[1]{{}}

\newcommand{\sgn}{\mathop{\rm sgn}}
\newcommand{\tr}{\mathop{\rm Tr}}
\newcommand{\rk}{\mathop{\rm rk}}
\newcommand{\rank}{\mathop{\rm rank}}
\newcommand{\corank}{\mathop{\rm corank}}
\newcommand{\range}{\mathop{\rm Range\,}}
\newcommand{\supp}{\mathop{\rm supp}}
\newcommand{\p}{\partial}
\renewcommand{\P}{\grave{\partial}}
\newcommand{\yDelta}{\grave{\Delta}}
\newcommand{\yD}{\grave{D}}
\newcommand{\yeurD}{\grave{\eur{D}}}
\newcommand{\yeubD}{\grave{\eub{D}}}
\newcommand{\at}[1]{\vert\sb{\sb{#1}}}
\newcommand{\At}[1]{\biggr\vert\sb{\sb{#1}}}
\newcommand{\vect}[1]{{\bold #1}}
\def\R{\mathbb{R}}
\newcommand{\C}{\mathbb{C}}
\def\hvar{{\hbar}}
\newcommand{\N}{\mathbb{N}}\newcommand{\Z}{\mathbb{Z}}
\newcommand{\Abs}[1]{\left\vert#1\right\vert}
\newcommand{\abs}[1]{\vert #1 \vert}
\newcommand{\Norm}[1]{\left\Vert #1 \right\Vert}
\newcommand{\norm}[1]{\Vert #1 \Vert}
\newcommand{\Const}{{C{\hskip -1.5pt}onst}\,}
\newcommand{\sothat}{{\rm ;}\ }
\newcommand{\Range}{\mathop{\rm Range}}
\newcommand{\ftc}[1]{$\blacktriangleright\!\!\blacktriangleright$\footnote{AC: #1}}



\newcommand{\const}{\mathop{\rm const}}

\renewcommand{\theequation}{\thesection.\arabic{equation}}

\makeatletter\@addtoreset{equation}{section}
\makeatother

\def\Tau{\mathcal{T}}

\def\os{{o}}
\def\ol{{O}}
\def\dist{\mathop{\rm dist}\nolimits}
\def\spec{\sigma}
\def\mod{\mathop{\rm mod}\nolimits}
\renewcommand{\Re}{\mathop{\rm{R\hskip -1pt e}}\nolimits}
\renewcommand{\Im}{\mathop{\rm{I\hskip -1pt m}}\nolimits}

\section{Introduction}

In the present paper we suggest the scenario, in which the recently discovered $125$ GeV h - boson \cite{CMSHiggs,ATLASHiggs} is  composite. According to the suggested model it is composed of all fermions of the Standard Model (SM). The scale of the hidden strong dynamics is supposed to be of the order of several TeV. We suppose, that the W and Z boson masses are determined by the condensate of the $125$ GeV $h$ - boson according to the Higgs mechanism \cite{Englert,Higgs}. All Dirac fermion masses are determined by the $h$ - boson as well. In the present paper we do not consider the Majorana masses at all and assume, that neutrinos have extremely small Dirac masses \cite{neutrino}. The low energy effective theory contains the four - fermion interaction \cite{Nambu} between all SM fermions. Our model differs from the conventional models with four - fermion interactions \cite{NJL}, in which the top quark is condensed. The model of top - quark condensation was first suggested in \cite{Miransky} (and developed later in \cite{Marciano:1989xd,topcolor1,Hill:1991at,Hill:2002ap,Miransky:1994vk}). The idea, that the Higgs boson may be composed of the known SM fermions was discussed even earlier, in \cite{Terazawa}, together with the certain preon models (however, in \cite{Terazawa} there was no emphasis in the dominating  role of the top quark and the compositeness of the Higgs boson was considered together with the compositeness of quarks and leptons). There are two main aspects, in which the model discussed in the present paper differs from the conventional models of top  - quark condensation:

\begin{enumerate}

\item{} The four - fermion interaction is  non - local. It contains the form - factors $G$ that correspond to the interaction between the composite Higgs boson and the pair fermion - antifermion. The formfactors  depend on three scalar parameters of the dimension of mass squared: $q^2$, $p^2$, $k^2$, where $p$ and $k$ are the 4 - momenta of the fermion and anti - fermion while $q=p-k$ is the 4 - momentum of the composite scalar boson. $G(p,k) = g(q^2,p^2,k^2)$  are different for different fermions at large distances (both time - like and space - like), i.e. if at least one of the quantities  $|q^2|,|p^2|,|k^2|$ is much smaller, than the Electroweak scale $M_Z \approx 90$ GeV. However, at small space - like and time  - like distances (i.e. for $|q^2| \sim |p^2| \sim |k^2| \sim M_Z^2$) those form - factors become equal for all SM fermions.
    Unlike for the other fermions the mentioned form - factor for the top - quark is assumed to be independent of momenta.

\item{} Our model with the four - fermion interaction is not renormalizable. Therefore, it is the effective theory only and its output strongly depends on the regularization scheme. Contrary to the approach of the mentioned above papers on the top  - quark condensation models we do not use the conventional cutoff regularization. We use zeta regularization  \cite{zeta,McKeon}. It allows to construct the effective theory without any ultraviolet divergencies.

\end{enumerate}

The $1/N_c$ expansion is a good approximation within our effective theory. This is because the dangerous ultraviolet divergences are absent. This is well - known, that these divergences break the $1/N_c$ expansion in the NJL models defined in ordinary cutoff regularization. For example, in \cite{cvetic} it is shown, that when the ultraviolet cutoff is much larger, than the generated masses, the next to leading order $1/N_c$ approximation to various quantities does not give the corrections  smaller, than the leading one. However, this occurs because of the terms that are formally divergent in the limit of infinite cutoff. Therefore, in the regularization, that is free from divergences the next to leading approximation is suppressed by the factor $1/N_c$ compared to the leading approximation. Moreover, for the calculation of various quantities related to the extra high energy processes (the processes with the characteristic energies much larger, than $m_t$) the next - to leading order approximation is suppressed by the factor $1/N_{\rm total}$, where $N_{\rm total} = 24$ is the total number of the Standard Model fermions (the color components are counted as well, so that we have $N_{\rm total} = 3\, {\rm generations}\, \times \, (3 \, {\rm colors} + 1\, {\rm lepton}) \, \times \, 2 \, {\rm isospin}\, {\rm components} = 24$).

Zeta regularization \cite{zeta,McKeon} provides the natural mechanism of cancelling the unphysical divergent contributions to various physical quantities in the field theories. For example, being applied to quantum hydrodynamics it provides the absence of the unphysical divergent contributions to vacuum energy due to the phonon loops. The latter are known to be cancelled exactly by the physics above the natural cutoff of the hydrodynamics \cite{quantum_hydro}. If zeta regularization is applied to quantum hydrodynamics, those divergences do not appear from the very beginning. The cancellation of the mentioned divergences in quantum hydrodynamics occurs due to the thermodynamical stability of vacuum \cite{quantum_hydro}. That's why zeta regularization is the natural way to incorporate the principle of the thermodynamical stability of vacuum already at the level of low energy approximation (i.e. at the level of the hydrodynamics).  It was suggested to apply the similar principles of stability to subtract divergences from the vacuum energy in the quantum theory in the presence of gravity \cite{hydro_gravity} and to subtract quadratic divergences from the contributions to the Higgs boson masses in the NJL models of composite Higgs bosons \cite{Z2013JHEP,VZ2013,VZ2012,VZ2013_2}. Now we understand, that the zeta regularization allows to implement such principles of stability to various theories in a very natural and common way. That's why this regularization may appear to be not only the regularization, but the essential ingredient of the quantum field theory.

The approach presented in the given paper allows to derive the relation $M_H^2 = m_t^2/2$ on the level of the leading order of the $1/N_c$ expansion. In the simplest models of the top - quark condensation the mass of the composite Higgs boson is typically related to the mass of the top quark by the relation $M_H^2 = 4 m^2_t$. The resulting mass $\approx 350$ GeV contradicts to the recent discovery of the   $125$ GeV $h$ - boson.  It is worth mentioning, that there were attempts to construct the theory that incorporates the existence of several composite Higgs bosons $M_{H,i}$ related to the top - quark mass by the  Nambu Sum rule $\sum M^2_{H,i} = 4 m_t^2$. Those models admit the appearance of the $125$ GeV $h$ - boson and  predict the existence of its Nambu partners \cite{VZ2012,VZ2013,VZ2013_2}. The approach of the present paper predicts the only composite Higgs boson with the observed mass around $m_t/\sqrt{2} \approx 125$ GeV.

The paper is organized as follows. In Section \ref{Sectfermions} we define the model under consideration.  In Section \ref{secteff} we present the effective action and evaluate physical quantities (the practical calculations using zeta regularization are placed in Appendix). In Section \ref{sectconcl} we end with the conclusions.

\section{The model under consideration}
\label{Sectfermions}

\subsection{Notations}

SM fermions carry the following indices:

\begin{enumerate}

\item{} Weil spinor indices. We denote them by large Latin letters $A,B,C,...$

\item{}  $SU(2)$ doublet indices (isospin) both for the right - handed and for the left - handed spinors are denoted by small Latin letters $a,b,c,...$ For the left - handed doublets index $a$ may be equal to $0$ and $1$. For the right - handed singlets of the Standard Model this index takes values $U, D, N, E$ that denote the up quark, down quark, neutrino and electron of the first generation and the similar states from the second and the third generations.

\item{} Color $SU(3)$ indices are denoted by small Latin letters $i,j,k,...$

\item{} The generation indices are denoted by bold small Latin letters $\bf a,b,c,...$

\end{enumerate}

Besides, space - time indices are denoted by small Greek letters $\mu, \nu, \rho, ...$
Let us introduce the following notations:
\begin{eqnarray}
&& L^{\bf 1} = \left(\begin{array}{c}u_L\\d_L\end{array}\right), \quad L^{\bf 2} = \left(\begin{array}{c}c_L\\s_L\end{array}\right), \quad L^{\bf 3} = \left(\begin{array}{c}t_L\\b_L\end{array}\right)\nonumber\\ && R_U^{\bf 1} = u_R, \quad R_U^{\bf 2} = c_R, \quad R_U^{\bf 3} = t_R
\nonumber\\ &&  R_D^{\bf 1} = d_R, \quad R_D^{\bf 2} = s_R, \quad R_D^{\bf 3} = b_R
\end{eqnarray}

and

\begin{eqnarray}
&& {\cal L}^{\bf 1} = \left(\begin{array}{c}\nu_L\\e_L\end{array}\right), \quad {\cal L}^{\bf 2} = \left(\begin{array}{c}\nu_{\mu L}\\\mu_L\end{array}\right), \quad {\cal L}^{\bf 3} = \left(\begin{array}{c}{\nu}_{\tau L}\\ \tau_L\end{array}\right)\nonumber\\ && {\cal R}_N^{\bf 1} = \nu_R, \quad {\cal R}_N^{\bf 2} = \nu_{\mu R}, \quad {\cal R}_N^{\bf 3} = \nu_{\tau R}
\nonumber\\ && {\cal R}_E^{\bf 1} = e_R, \quad {\cal R}_E^{\bf 2} = \mu_R, \quad {\cal R}_E^{\bf 3} = \tau_R
\end{eqnarray}

\subsection{The action}

In our consideration we neglect gauge fields. The partition function has the form:
\begin{equation}
Z = \int D \bar{\psi}D\psi e^{iS},
\end{equation}
By $\psi$ we denote the set of all Standard Model fermions. The action $S =  \int d^4x \bar{\psi} i \partial \gamma  \psi + S_I$ contains two terms. The first one is the kinetic term.
The second one is related to the hidden strong interaction between the SM fermions, that results in the appearance of the additional four - fermion non - local interaction term
\begin{widetext}
\begin{eqnarray}
S_I &=&\frac{1}{  M_I^2} \sum_{\bf a}\int d^4x_1 d^4 x_2 d^4 y_1 d^4 y_2 d^4 z \Bigl(\bar{L}_b^{{\bf a}}(x_1) R^{\bf a}_U(x_2) G^{{\bf a}}_U(x_1-z, x_2-z)  + \bar{\cal L}_b^{\bf a}(x_1) {\cal R}^{\bf a}_N(x_2) G^{{\bf a}}_N(x_1-z,x_2-z)\nonumber\\&& + \bar{R}^{\bf a}_D(x_1) {L}_c^{{\bf a}}(x_2)\epsilon_{bc}G^{{\bf a}}_D(x_1-z,x_2-z) + \bar{\cal R}^{\bf a}_E(x_1) {\cal L}_c^{\bf a}(x_2) \epsilon_{bc}G^{{\bf a}}_E(x_1-z,x_2-z) \Bigr)\nonumber\\&&
 \Bigl(\bar{G}^{{\bf a}}_D(y_2-z,y_1-z)\bar{L}_d^{{\bf a}}(y_1) R^{\bf a}_D(y_2)\epsilon_{bd}  + \bar{G}^{{\bf a}}_E(y_2-z,y_1-z)\bar{\cal L}_d^{\bf a}(y_1) {\cal R}^{\bf a}_E(y_2) \epsilon_{bd}\nonumber\\&& + \bar{G}^{{\bf a}}_U(y_2-z,y_1-z)\bar{R}^{\bf a}_U(y_1) {L}_b^{{\bf a}}(y_2) + \bar{G}^{{\bf a}}_N(y_2-z,y_1-z)\bar{\cal R}^{\bf a}_N(y_1) {\cal L}_b^{\bf a}(y_2)  \Bigr),
 \label{I4}
\end{eqnarray}
\end{widetext}
Here $M^2_I$ is the dimensional parameter (it will be shown below that it's bare value is negative and is around $M_I^2 \approx - M_Z^2 \approx-[90 {\rm GeV}]^2$). Functions  $G^{\bf a}_{a}(y_1-z,y_2-z)$ for $a = U,D,N,E$ are the form - factors mentioned in the introduction defined in coordinate space ($\bar{G}^{\bf a}_a$ means complex conjugation, $\epsilon_{ab}$ is defined in such a way, that $\epsilon_{12} = - \epsilon_{21} = 1$).  In momentum representation we have
\begin{widetext}
\begin{eqnarray}
S_I &=&\frac{V}{M_I^2} \sum_{\bf a} \sum_{q = p-k = p^{\prime} - k^{\prime}} \Bigl(\bar{L}_b^{{\bf a}}(p) R^{\bf a}_U(k) G^{{\bf a}}_U(p,k)  + \bar{\cal L}_b^{\bf a}(p) {\cal R}^{\bf a}_N(k) G^{{\bf a}}_N(p,k)\nonumber\\&& + \bar{R}^{\bf a}_D(p) {L}_c^{{\bf a}}(k)\epsilon_{bc}G^{{\bf a}}_D(p,k) + \bar{\cal R}^{\bf a}_E(p) {\cal L}_c^{\bf a}(k) \epsilon_{bc}G^{{\bf a}}_E(p,k) \Bigr)\nonumber\\&&
 \Bigl(\bar{G}^{{\bf a}}_D(p^{\prime},k^{\prime})\bar{L}_d^{{\bf a}}(k^{\prime}) R^{\bf a}_D(p^{\prime})\epsilon_{bd}  + \bar{G}^{{\bf a}}_E(p^{\prime},k^{\prime})\bar{\cal L}_d^{\bf a}(k^{\prime}) {\cal R}^{\bf a}_E(p^{\prime}) \epsilon_{bd}\nonumber\\&& + \bar{G}^{{\bf a}}_U(p^{\prime},k^{\prime})\bar{R}^{\bf a}_U(k^{\prime}) {L}_b^{{\bf a}}(p^{\prime}) + \bar{G}^{{\bf a}}_N(p^{\prime},k^{\prime})\bar{\cal R}^{\bf a}_N(k^{\prime}) {\cal L}_b^{\bf a}(p^{\prime})  \Bigr), \label{IM}
\end{eqnarray}
\end{widetext}
Here $V$ is the four - volume. Fermion fields in momentum space are denoted as $\psi(q) = \frac{1}{V}\int d^4x e^{-i q x} \psi(x)$ while the form - factors in momentum space are defined as $G^{\bf a}_{a}(p,k) = \int d^4x d^4 y e^{-i p x + i k y} G^{\bf a}_{a}(x,y)$ for $a = U,D,N,E$. According to our supposition all functions
$G^{\bf a}_{a}(p,k) = g^{\bf a}_{a}(q^2,p^2,k^2)$ vary between fixed values much smaller than unity at small momenta and unity at large values of $|q^2|,|p^2|,|k^2|$:
\begin{eqnarray}
{G}^{\bf a}_{a}(p,k) &=& g^{\bf a}_{a}(q^2,p^2,k^2)  \rightarrow 1 \nonumber\\&& (|q^2|,|p^2|,|k^2| \sim M_Z^2 \sim [90\,{\rm GeV}]^2 ); \nonumber\\ {G}^{\bf a}_{a}(p,k) & = & g^{\bf a}_{a}(q^2,p^2,k^2) \rightarrow \kappa^{\bf a}_{a}\nonumber\\&& (|q^2|,|p^2|,|k^2| \ll M_Z^2)\nonumber\\ |{G}^{\bf a}_{a}(p,k)| & = & |g^{\bf a}_{a}(q^2,p^2,k^2)| \ll 1\nonumber\\&& (|q^2| \sim M_Z^2;\, |p^2|,|k^2| \ll M_Z^2;\nonumber\\&& {\rm except}\, {\rm for} \,{\rm top}\, {\rm quark})\label{condition}
\end{eqnarray}
Here $M_Z \approx 90$ GeV is the Electroweak scale,  $\kappa^{\bf a}_{a}$ are the dimensionless coupling constants that are all much smaller than unity except for the top quark. Because of the isotropy of space - time all $G^{\bf a}_{a}$ are the functions of the invariants $p^2, k^2$, and $q^2 = (p-k)^2$ only. Besides, in order to provide the left - right symmetric mass matrix (see below) we require $G^{\bf a}_a(x,y) = \bar{G}^{\bf a}_a(y,x)$ that leads to $G(p,k) = \bar{G}(k,p)$ and to $g^{\bf a}_{a}(q^2,p^2,k^2) = \bar{g}^{\bf a}_{a}(q^2,k^2,p^2)$. In principle, we may consider the real - valued form - factors $G(x,y)=G(y,x)$ that would result in $G^{\bf a}_{a}(p,k) = \int d^4x d^4 y e^{-i p x + i k y} G^{\bf a}_{a}(x,y) = \int d^4x d^4 y e^{-i p y + i k x} G^{\bf a}_{a}(x,y) = \bar{G}^{\bf a}_a(k,p) = G^{\bf a}_a(-k,-p)$. Since $G(k,p)=G(-k,-p) = g((p-k)^2,k^2,p^2)$ we would arrive at the real - valued form - factors in momentum space.  We imply that Eq. (\ref{condition}) is valid for the complex - valued $p^2,q^2,k^2$ in the upper half of the complex plane.   Besides, we require, that
for the the top quark
\begin{equation}
{G}_{t}(p,k)\equiv {G}^{\bf 3}_{U}(p,k) \approx 1, \label{vtop}
\end{equation}
for any $p,k$. We may choose, for example,
\begin{widetext}
\begin{equation}
g^{\bf a}_a(q^2,p^2,k^2) = \frac{q^4 + \alpha^{\bf a}_{a} M_I^4}{q^4 + \gamma^{\bf a}_a M_I^4}\times \frac{p^4 + \beta^{\bf a}_{a} M_I^4}{p^4 + \lambda^{\bf a}_a M_I^4}\times  \frac{k^4 + \beta^{\bf a}_{a} M_I^4}{k^4 + \lambda^{\bf a}_a M_I^4}\label{ex2}
\end{equation}
\end{widetext}
with some constants $\alpha, \beta, \gamma, \lambda$ such that $\frac{\alpha^{\bf a}_{a} \beta^{\bf a}_{a} \beta^{\bf a}_{a}}{\gamma^{\bf a}_{a}\lambda^{\bf a}_{a}\lambda^{\bf a}_{a}} = \kappa^{\bf a}_{a}$ while $\alpha^{\bf a}_{a}, \beta^{\bf a}_{a},\gamma^{\bf a}_{a},\lambda^{\bf a}_{a}\ll 1$, and $\alpha^{\bf 3}_{U}= \beta^{\bf 3}_{U}=\gamma^{\bf 3}_{U}=\lambda^{\bf 3}_{U}=1$.
We may also interpolate the Form - factors as follows (for all fermions except for the top - quark):
\begin{equation}
 g^{\bf a}_a(q^2,p^2,k^2) = \left\{\begin{array}{c}1 \, {\rm for} \, |q^2|,|p^2|,|k^2| > M^2_0\\
\kappa^{\bf a}_a << 1  \, {\rm otherwise}\end{array}\right| \label{ex1}
\end{equation}
with a dimensional parameter $M_0$ such that $ M_0 \ll M_Z$ (where $M_Z$ is the $Z$ - boson mass). For the top - quark we set  ${G}^{\bf 3}_{U}(p,k) =  g^{\bf 3}_U(q^2,p^2,k^2) = \kappa^{\bf 3}_{U} =1$ .
 Let us introduce the auxiliary scalar $SU(2)$ doublet $H$. The resulting action receives the form
 \begin{widetext}
\begin{eqnarray}
S & = &  \int d^4x\Bigl[ \bar{\psi} i \partial \gamma \psi -  |H|^2 M^2_I\bigr]
\nonumber\\&& - \sum_{{\bf a}}\int d^4 x_1 d^4x_2 d^4 z \Bigl[\Bigl(\Bigl[\bar{L}_b^{{\bf a}}(x_1) R^{\bf a}_U(x_2)  {G}^{\bf a}_{U}(x_1-z,x_2-z) H^b(z)  + \bar{\cal L}_b^{\bf a}(x_1) {\cal R}^{\bf a}_N (x_2) {G}^{\bf a}_{N}(x_1-z,x_2-z) H^b(z) \Bigr]\nonumber\\&& + \Bigl[\bar{L}_c^{{\bf a}}(x_1) R^{\bf a}_D (x_2)  \bar{G}^{\bf a}_{D}(x_2-z,x_1-z) \bar{H}^b(z)\epsilon_{bc} + \bar{\cal L}_c^{\bf a}(x_1) {\cal R}^{\bf a}_E(x_2) \bar{G}^{\bf a}_{E}(x_2-z,x_1-z) \bar{H}^b(z) \epsilon_{bc} \Bigr]+(h.c.) \Bigr) \Bigr] \nonumber \label{Sf2222}
\end{eqnarray}
\end{widetext}
In momentum space we have:
\begin{widetext}
\begin{eqnarray}
S & = & V \sum_p \Bigl[ \bar{\psi}(p) p \gamma \psi(p) -  H^+(p)H(p) M^2_I\Bigr]
\nonumber\\&& - V\sum_{{\bf a}}\sum_{q=p-k}\Bigl[\bar{L}_b^{{\bf a}}(p) R^{\bf a}_U(k)  {G}^{\bf a}_{U}(p,k) H^b(q) + \bar{\cal L}_b^{\bf a}(p) {\cal R}^{\bf a}_N(k)  {G}^{\bf a}_{N}(p,k) H^b(q)  +(h.c.)\Bigr]\nonumber\\&& -V\sum_{{\bf a}}\sum_{q=p-k} \Bigl[\bar{L}_c^{{\bf a}}(k) R^{\bf a}_D(p)  \bar{G}^{\bf a}_{D}(p,k) \bar{H}^b(q)\epsilon_{bc} + \bar{\cal L}_c^{\bf a}(k) {\cal R}^{\bf a}_E(p) \bar{G}^{\bf a}_{E}(p,k)\bar{H}^b(q)\epsilon_{bc} +(h.c.)\Bigr]  \Bigr],  \label{Sf2222M}
\end{eqnarray}
\end{widetext}
where $H(q) = \frac{1}{V}\int d^4 x H(x) e^{-iqx}$.

The important property of the functions $g^{\bf a}_a(q^2,p^2,k^2)$ is that the Wick rotation from the positive values of $p^2,q^2,k^2$ to negative values of $p^2,q^2,k^2$ (i.e. positive values of Euclidean momenta squared) keeps the relation
\begin{eqnarray}
&& {G}^{\bf a}_{a}(p,k)  =  g^{\bf a}_{a}(q^2,p^2,k^2)  \rightarrow 1 \nonumber\\&& (|q^2|,|p^2|,|k^2| \sim M_Z^2 \sim [90\,{\rm GeV}]^2 )\label{cond0}
\end{eqnarray}
 This will allow us to evaluate loop integrals using the Euclidean expressons (see Appendix).
 It is worth mentioning, that the invariance of Eq.(\ref{cond0}) under the Wick rotation is seen directly in the form of the Form - factors of Eq. (\ref{ex2}).

\subsection{Fermion masses}

In the following we fix Unitary gauge $H = (v + h, 0)^T$, where $v$ is vacuum average of the scalar field. We denote the Dirac $4$ - component spinors in this gauge corresponding to the SM fermions by $\psi^{\bf a}_{a}$. We omit angle degrees of freedom to be eaten by the gauge bosons.
In this gauge the propagator $Q^{\bf a}_{a}$ , $a = U,D,N,E$ of the SM fermion has the form:
\begin{equation}
[Q^{\bf a}_{a}]^{-1} = \hat{p} \gamma - v G^{\bf a}_{a}\,(p,p)
\end{equation}
For each fermion except for the top - quark there is the pole of this propagator at $|p^2| \ll M_Z^2$ that gives the fermion mass
\begin{equation}
m^{\bf a}_{a} \approx v G^{\bf a}_{a}\,(0,0) \approx v \kappa^{\bf a}_{a}
\end{equation}
For the top quark we have $m^{\bf 3}_U = v$.
The value of $v$ is to be calculated using the gap equation that is  the extremum condition for the effective action as a function of $h$.
The effective action of Eq. (\ref{Sf2222}) can be rewritten as
\begin{eqnarray}
S &=&  \int d^4x\Bigl( \bar{\psi}(x)( i \partial \gamma - M  ) \psi(x) - \bar{\psi}(x)\Bigl[ \hat{G}_h \psi\Bigr](x)\nonumber\\&&- M_I^2 (v+h(x))^2  \Bigr),\label{Smixed}
\end{eqnarray}
where $M$ is the mass matrix. It is diagonal, with the diagonal  components $m^{\bf a}_{a}$ ($a = U,D,N,E; {\bf a} = {\bf 1,2,3}$) given by $\kappa^{\bf a}_{a} v$. We use here the following notation for the h - dependent operator $\hat{G}_h$:
\begin{equation}
[\hat{G}_h\xi](x_1) = \int d^4 z d^4 x_2 \,  \xi(x_2)  G(x_1-z,x_2-z) h(z) \label{conv}
\end{equation}
This operator is hermitian as follows from the condition $G(x,y) = \bar{G}(y,x)$.  
We imply, that the $h$ - boson of Eq. (\ref{Sf2222}) is responsible for the formation of masses of $W$ and $Z$ bosons. Here and below we omit indices ${\bf a}, {a}$ for $G$ and $\kappa$. The sum over these indices is implied in the following expressions. $G$ and $\kappa$ are considered as the diagonal matrices with the diagonal elements $G^{\bf a}_{a}$ and $\kappa^{\bf a}_{a}$ correspondingly.

The interaction at the momenta of fermions $|p^2|, |k^2| \sim [90\,{\rm GeV}]^2$ and the momentum of the Higgs boson $|q^2| = |(p-k)^2| \sim [90\, {\rm GeV}]^2$  gives rise to the $U(N_{\rm total})$ (with $N_{\rm total} = 24$) symmetric interaction between the real - valued Higgs field excitations $h$ and the SM fermions. The interaction at momenta $|(p-k)^2| \approx 0$ corresponds to the Higgs field condensate. The existing poles of fermion propagators give the masses of all SM quarks and leptons.  The value of $v$ is to be determined through the requirement, that $\frac{\delta}{\delta h} S_{\rm eff}(h) = 0$ at $h = 0$, where $S_{\rm eff}$ is the  effective action (obtained after the integration over the fermions).

\section{Effective low energy action and the evaluation of physical quantities.}
\label{secteff}

\subsection{Effective action for the $h$ - boson. Gap equation and Higgs boson mass.}

The effective action for the theory with action Eq. (\ref{Smixed}) as a function of the field $h$ is obtained as a result of the integration over fermions. In zeta regularization this effective action is calculated in Appendix up to the terms quadratic in $h$. This corresponds to the leading $1/N_c$ approximation.

The total one - loop  effective action for the $h$ - boson receives the form:
\begin{widetext}
\begin{equation}
S[h] =  \int {d^4x} \Bigl[- M_I^2 (v+h)^2 -  \hat{C} v^2 (v+h)^2 +  h(x)\, Z_h^2\, (w({\hat p}^2)\hat{p}^2 - M_H^2) h(x) \Bigr],\label{Sh}
\end{equation}
\end{widetext}
where
\begin{eqnarray}
&&  w(p^2) \approx 1, \quad (|p^2| \sim M_H^2) \nonumber\\
&&  w(p^2) \approx 1/8, \quad (|p^2| \ll M_H^2) \nonumber\\ && Z_h^2  \approx   \frac{N_{\rm total}}{16 \pi^2}{\rm log}\, \frac{\mu^2}{m_t^2}\nonumber\\&& M_H^2 = 4 N_c m_t^2/N_{\rm total};\nonumber\\&& \hat{C} = \frac{N_c}{8 \pi^2 } {\rm log}\frac{\mu^2}{m_t^2} \label{zw}
\end{eqnarray}
According to our supposition $\kappa_t = 1$, and the mass of the top - quark is given  by $v = m_t$. Here we neglect the imaginary part of the effective action that is much smaller, than the real part for the momenta squared of the h - boson smaller, than $4 m_t^2$. For $p^2 \ge 4 m_t^2$ the decay of the h - boson into the pair $t\bar{t}$ should be taken into account through the imaginary part of effective action.

Recall, that the resulting expressions for various quantities in zeta - regularization contain the scale parameter $\mu$. We identify this parameter with the typical scale of the interaction that is responsible for the formation of the composite Higgs boson. Below it will be shown, that the value of $\mu$ consistent with the observed masses of Higgs, W, and Z - bosons is $\mu \approx 5$ TeV.

The vacuum value $v$ of $H$ satisfies gap equation
\begin{equation}
\frac{\delta}{\delta h} S[h] = 0
\end{equation}
 Then we get the negative value for the bare mass parameter of the four - fermion interaction:
\begin{equation}
M_I^2  = -\frac{N_c}{8 \pi^2 } {m_t^2}\, {\rm log}\frac{\mu^2}{m_t^2}
\end{equation}
This negative value $M_I^2 \approx - M_Z^2$ means, that the bare four - fermion interaction at the high energy scale $\mu \gg m_t$ of Eq. (\ref{I4}) is repulsive. However, the appearance of the vacuum average means, that this repulsive interaction is subject to finite renormalization: at low energies $\ll \mu$, where the Higgs boson mass is formed, the renormalized interaction between the fermions is attractive. One can see, that there is only one pole of the propagator for the field $h$. As a result the Higgs mass is indeed given by
\begin{eqnarray}
&& M^2_H \approx 4 N_c m_t^2/N_{\rm total}  = m^2_t/2 \approx 125 \, {\rm GeV}\nonumber
\end{eqnarray}

\subsection{W and Z boson masses. The new interaction scale $\mu$.}
\label{sectmu}

We may consider Eq. (\ref{Sh}) as the low energy effective action at most quadratic in the scalar field. There exists the effective action written in terms of the field $H$. It should contain the $\lambda |H|^4$ term in order to provide the nonzero vacuum average $\langle H \rangle = (v,0)^T$. Comparing the coefficients at the different terms of this effective action with that of Eq. (\ref{Sh}) we arrive at
\begin{widetext}
\begin{equation}
S[H] =  \int {d^4x}  \Bigl(  H^+(x)\, Z_h^2 w(-D^2)\,\Bigl(-D^2\Bigr) \,H(x) - \frac{Z_h^2}{8}(|H|^2 - v^2)^2 \Bigr)\label{SH}
\end{equation}
\end{widetext}
Here the gauge fields of the SM are taken into account. As a result the usual derivative $\partial$ of the field $H$ is  substituted by the covariant one $D = \partial + i B$ with $B = B^a_{SU(2)} \sigma^a + Y B_{U(1)}$ (where $Y = -1$ is the hypercharge of $H$). Functions $w, Z_h$ are defined in Appendix. They depend strongly on the particular choice of the form - factors $G^{\bf a}_a$ and satisfy the conditions of Eq. (\ref{zw}).

In order to calculate gauge boson masses we need to substitute into Eq. (\ref{SH}) the $SU(2)\otimes U(1)$ gauge field $B$ instead of $i D$ and $(v,0)^T$ instead of $H$. As a result we obtain the effective potential for the field $B$:
\begin{equation}
V_B \approx   m_t^2  \,Z^2_h w(\|B\|^2)\, \|B\|^2 \label{VB}
\end{equation}
Here we denote $\|B\|^2 = (1,0)B^2(1,0)^T = \Bigl((B^3_{SU(2)} - B_{U(1)})^2 + [B_{SU(2)}^1]^2 + [B_{SU(2)}^2]^2\Bigr)$.  Up to the terms quadratic in $B$ we get:
\begin{eqnarray}
V^{(2)}_B &\approx & \frac{\eta^2}{2}\, \|B\|^2 \label{eta0}
\end{eqnarray}

The renormalized vacuum average of the scalar field $\eta$ should be equal to $\approx 246$ GeV in order to provide the appropriate values of the gauge boson masses. In order to evaluate the gauge boson masses in the given model we cannot neglect nontrivial dependence of  $w(\|B\|^2)$ on $B$ in Eq. (\ref{VB}). To demonstrate this let us evaluate the typical value of $B$ that enters Eq. (\ref{VB}). Our
calculation follows the classical method for the calculation of fluctuations in quantum field theory and statistical theory (see, for example, volume 5, paragraph 146 of \cite{Landau}).
 The typical value of $B$ is given by the average fluctuation $\langle \|\bar{B}\|^2\rangle$ within the four - volume $\Omega$ of the linear size $\sim \frac{1}{M_Z}$, where $M_Z$ is the Z - boson mass, while $\|\bar{B}\|^2 = \frac{1}{|\Omega|}\int_{\Omega} \|B\|^2 d^4x$. The direct estimate gives  $\langle \|B\|^2\rangle \approx -\frac{\partial}{\partial |\Omega| M^2_Z}\, {\rm log} \, \int dB e^{-|\Omega| M^2_Z \|B\|^2 } = \frac{1}{|\Omega|M^2_Z}\sim M^2_Z$. (We calculate the integral in space - time of Euclidean signature and then rotate back the final result.) As a result in Eq. (\ref{VB}) we may substitute $w(M_Z^2)\approx 1$ instead of $w(\|B\|^2)$. This gives
\begin{eqnarray}
S_B &\approx &\frac{\eta^2}{2}\, \|B\|^2 \approx  \frac{ N_{\rm total}}{16 \pi^2} m_t^2\,  {\rm log}\, \frac{\mu^2}{m_t^2} \, \|B\|^2\nonumber\\ && \|B\|^2 \sim  M^2_Z \approx [90 \, {\rm GeV}]^2
\end{eqnarray}
That's why we arrive at the following expression for $\eta$
\begin{equation}
{\eta^2} \approx 2  Z_h^2 v^2 = \frac{2 N_{\rm total}}{16 \pi^2} m_t^2 {\rm log}\, \frac{\mu^2}{m_t^2}\label{eta_}
\end{equation}
From here we obtain $\mu \sim 5$ TeV.

\subsection{Branching ratios and the Higgs production cross - sections}

The Higgs boson production cross - sections and the decays of the Higgs bosons are described by the effective lagrangian
\cite{status}:
\begin{widetext}
\bea
\label{eq:1}
L_{eff}  &=  &
   {2 m_W^2  \over \eta}  h  \,   W_\mu^+ W_\mu^-  +     {m_Z^2 \over \eta} h  \,  Z_\mu  Z_\mu
 + c^{}_{g}  {\alpha_s \over 12 \pi \eta} h \, G_{\mu \nu}^a G_{\mu \nu}^a \nonumber  \\&& +  c^{}_{\gamma} { \alpha \over \pi \eta} h \, A_{\mu \nu} A_{\mu \nu} -   c_b {m_b \over \eta } h \,\bar b b  -  c_c  {m_c \over \eta}  h \,   \bar c c \ -   c_{\tau} {m_{\tau} \over \eta } h \,\bar \tau \tau .
\eea
\end{widetext}
Here $G_{\mu\nu}$ and $A_{\mu\nu}$ are the field strengths of gluon and photon fields, $\eta$ the conventionally normalized vacuum average of the scalar field $\eta \approx 246\, {\rm GeV}$.
This effective lagrangian should be considered at the tree level only and describes the channels $h \rightarrow gg, \gamma \gamma, ZZ, WW, \bar b b, \bar c c, \bar \tau \tau$. The fermions and $W$ bosons have been integrated out in the terms corresponding to the decays $h\rightarrow \gamma \gamma, gg$, and their effects are included in the effective  couplings $c_g$ and $c_\gamma$ (the similar constants in the SM are given by $c_{g}   \simeq 1.03\,, c_{\gamma} \approx  -0.81$, see \cite{status}).
The effective lagrangian Eq. (\ref{eq:1}) is similar to that of the Standard Model. This results from our initial requirement, that the  Form - factors ${G}^{\bf a}_{a}(p,k)   \ll 1 $ for $|(p-k)^2| \sim M_Z^2;\, |p^2|,|k^2| \ll M_Z^2$ for all fermions except for the top quark. This reduces considerably the rates of the direct decays of the Higgs boson to light fermions.
However, unlike the SM the values of $c_{b}, c_c, c_{\tau}$ may differ from unity and are given by
\begin{eqnarray}
c_{b} &\approx & g(M_H^2, m_b^2, m_b^2)/\kappa^{\bf 3}_D,\nonumber\\ c_{c} & \approx & g(M_H^2, m_c^2, m_c^2)/\kappa^{\bf 2}_U\nonumber\\ c_{b} &\approx & g(M_H^2, m_{\tau}^2, m_{\tau}^2)/\kappa^{\bf 3}_E
\end{eqnarray}
The similar decay constants may be defined for all SM fermions: $c_{f^{\bf a}_a} \approx g(M_H^2, m_f^2, m_f^2)/\kappa^{\bf a}_a$. We assume, that these constants are not much larger than unity, so that the decays into the heavy fermions dominate like in the SM.

The consideration of constants $c_g$ and $c_{\gamma}$ is more involved. Those constants contain the fermion loops. Let us consider for the definiteness the constant $c_g$ (the consideration of $c_{\gamma}$ is similar). The contribution of each fermion (other than the top - quark) is given by the loop integral. The typical value of momentum circulating within the loop is of the order of the largest of the two external parameters: $M_H$ and $m_f$. As a result we cannot neglect the form - factor $G(p,k) \sim g(M_H^2,M_H^2,M_H^2)$ for the light fermions compared to that of the top quark. Within this integral there are three fermion propagators and two vertices proportional to $\gamma^{\mu}$. We arrive at the expression that is proportional to $m_f$ that results in $\delta c_g^{(f)} = \frac{2m_f}{M_H} \delta r_g^{(f)}$. We may estimate quantity $\delta r_g^{(f)}$ as follows. It is given by the dimensionless loop integral. For the definiteness let us consider the expression for the Form - factors given by Eq. (\ref{ex1}). We insert $G(p,k) \sim g(M_H^2,M_H^2,M_H^2) \approx 1$ into expression for the loop integral. However, while evaluating this loop integral we are to consider the region of momenta squared bounded from below by $M_0^2$. Besides, we should omit the contribution of the part of this expression related to the decay of the Higgs boson into the intermediate state composed of two light fermions. Such a contribution is related to the imaginary part of the fermion propagator proportional to $\delta(p^2 - m_f^2)$. Therefore, during the evaluation of such a contribution related to the intermediate fermion states on mass shell one should substitute the value of the form - factor $g(M_H^2, m_f^2, m_f^2) \ll 1$ instead of $g(M_H^2, M_H^2, M_H^2) \approx  1$. Thus, this contribution is suppressed for the light fermions. The value $g(M_H^2, M_H^2, M_H^2) \approx 1$ is to be substituted into the remaining (the main) part of the amplitude, and we should omit the imaginary - valued expressions during the calculation. For the rough evaluation of $r_g^{(f)}$ we use the expression for $A_f(\tau)$ given in Eq. (2.5) of \cite{status}. We substitute into the expression the value $\tau_0 = \frac{M^2_H}{4 M_0^2}$ instead of $\tau = \frac{M^2_H}{4 m_f^2}$ and multiply the resulting expression by the two factors: $\frac{m_t}{M_0} \approx \frac{M_H \sqrt{2}}{M_0} = 2 \sqrt{2}\, \sqrt{\tau_0}$ (takes into account the vertex $g(M_H^2,M_H^2,M_H^2) \approx 1$ instead of $M_0/m_t$) and $\sqrt{\tau_0}$ (takes into account that we deal with $\delta r_g^{(f)}$ instead of $\delta c_g^{(f)}$):
\begin{eqnarray}
\delta r^{(f)}_g &\approx& \frac{3 \sqrt{2} }{ \tau_0}\Bigl(\tau_0 - \frac{\tau_0 -1}{4}\Bigl[{\rm log} \Bigl|\frac{ \sqrt{1-1/\tau_0}+1}{ \sqrt{1-1/\tau_0}-1}\Bigr| \Bigr]^2 \Bigr)\label{rg}
\end{eqnarray}
As it was explained above, we omitted the imaginary part of the logarithm. One can see, that the value $\delta r_g = \delta r^{(f)}_g$ calculated in this way does not depend on the fermion flavor. This value ranges between $\sim -20$ at $M_0 = 10$ GeV and $\sim 0$ at $M_0 = 35$ GeV. We should notice that this is very rough evaluation that should be considered as the order of magnitude estimate only. The output from this estimate is that the expression for the quantity $r_g$ depends strongly on the particular form of the Form - factors, is negative and is of the order of $\sim -10$. Nevertheless, we feel this instructive to give the estimate for $c_g$ using the particular form of $r_g$ of Eq. (\ref{rg}). Namely, the contributions of the top - quark, bottom quark and charm quark dominate. The ratio $c_g/c^{(SM)}_g$ (where $c^{(SM)}_g$ is the SM value)  ranges from $\sim -1.0$ for $M_0 = 10$ GeV to $\sim +1.0$ for $M_0 = 35$ GeV. We conclude, therefore, that the contribution of the light fermions depends strongly on the particular form of the From - factors $g^{\bf a}_a$. However, there exists the possibility, that the value of the gluon fusion cross - section matches the present experimental constraints (see Fig. 12 of \cite{ATLASHIGGS} and Fig. 47 of \cite{CMSHIGGS}). In particular, the values $ c_g/c_g^{(SM)}  \sim \pm 1.0$ give the same cross - section for the process $gg \rightarrow h$ as the Standard Model.

The contribution of the light fermions to the decay constant $c_{\gamma}$ may be roughly estimated in the similar way. Now, the $b,c$ - quarks and $\tau$ - lepton dominate, and their contribution is given by $\delta c_{\gamma}^{(b,c,\tau)} = \Bigl(3 (1/3)^2 m_b + 3(2/3)^2 m_c + m_{\tau}\Bigr)\frac{2}{6 M_H} \delta r_g$. In the Standard Model value  $c^{(SM)}_{\gamma}$ the contribution of the $W$ - boson dominates. The modeling expression for $r_g$ of Eq. (\ref{rg}) gives the value of $c_{\gamma}/c^{(SM)}_{\gamma}$ that ranges from $\sim 1.4$ for $M_0 = 10$ GeV to $\sim 1.0$ for $M_0 = 35$ GeV. One can see, that the constant $c_{\gamma}$ also depends on the particular form of the Form - factors. There certainly exists the choice of the form - factors such that the resulting expressions for $c_g$, $c_{\gamma}$ match the present experimental constraints given by Fig. 12 of \cite{ATLASHIGGS} and Fig. 47 of \cite{CMSHIGGS}. There is a hint in the present experimental results. The best fit to the ratio $|c_{\gamma}/c^{(SM)}_{\gamma}|$ reported by ATLAS is $\sim 1.2$ while the best fit reported by CMS is $\sim 1.4$. At the same time our rough estimate described above gives this ratio within the interval $ (1.0, 1.4)$. The  best fit to the ratio $|c_g/c_g^{(SM)}|$ reported by ATLAS is $\sim 1.0$, the best fit reported by CMS is $\sim 0.7$ while our estimate gives this ratio within the interval $(0.0, 1.0)$.

We conclude, that in the model suggested here the branching ratios of Higgs decay into the light fermions, the branching ratio for the decay into two photons and the production cross - section for the process that goes through the gluon fusion $gg \rightarrow h$ depend strongly on the particular form  of the Form - factors of Eq. (\ref{IM}). In the two latter cases the dependence on the particular form of the Form - factors and the potential deviation from the SM value is the most dramatic. Nevertheless, in this subsection we demonstrated, that the values of the branching ratios and production cross - sections can be made matching the present experimental constraints (given, for example, in \cite{ATLASHIGGS} and \cite{CMSHIGGS}) by a certain choice of the Form - factors.

\section{Conclusions and discussion}
\label{sectconcl}

In this paper we suggest, that there exists the hidden strong dynamics behind the formation of the $125$ GeV Higgs boson. According to our scenario it is composed of all known SM fermions. The corresponding interaction has the form of the non - local four - fermion term of Eqs. (\ref{I4}), (\ref{IM}). The scale of the given interaction between the SM fermions is of the order of $\mu \approx 5$ TeV. The form - factors entering this term of the action depend on the values of the fermion momenta $p,k$.  At large values $|(p-k)^2|,|p^2|,|k^2| \sim [90\, {\rm GeV}]^2$ all form - factors are equal to unity, and the interaction with real - valued excitation of the Higgs boson becomes identical for all SM fermions. In the opposite limit of small momenta the symmetry is lost, and the form - factors are different for different fermions. This allows to obtain the observed hierarchy of masses for quarks and leptons. The Higgs production cross - section and the branching ratios of the Higgs boson decay depend strongly on the particular form of the Form - factors. The most dramatic dependence is encoded in the constants $c_g,c_{\gamma}$ of effective lagrangian Eq. (\ref{eq:1}) which are very sensitive to the new physics. These constants are related to the cross section of the Higgs boson production that goes through the gluon fusion and the branching ratio for the decay of the Higgs boson into two photons. A certain choice of the Form - factors gives the observable branching ratios and production cross - sections that do not deviate significantly from the SM values, and the present model matches existing experimental constraints. The further investigation of the LHC data may give more information and  more strong constraints on the experimentally allowed dependence of the Form - factors $g^{\bf a}_a$ of Eq. (\ref{IM}) on the values of momenta.

In our model the condensate of the composite $h$ - boson is responsible for the fermion masses. This distinguishes it from the ones, in which the $125$ GeV $h$ - boson is responsible for the masses of W and Z while there are the sources of the fermion masses different from the $h$ - boson (see, for example, \cite{Hashimoto:2009ty,Hashimoto:2009xi,Miransky2013}).
It is worth mentioning, that in the usual model of top - quark condensation \cite{Miransky},  the interaction scale was extremely high $\Lambda \sim 10^{15}$ GeV while in our case the interaction scale is $\mu \sim 5$ TeV.

In order to define the effective theory with the interaction term of Eqs. (\ref{I4}), (\ref{IM}) we use zeta regularization that gives the finite expressions for the considered observables. Bare interaction term at the scale of $\mu$ is repulsive. However, the loop corrections provide the appearance of the condensate for the composite scalar field. This distinguishes our effective theory defined using zeta regularization from the models with the four - fermion interaction defined using the usual cutoff regularization. Gap equation appears as an extremum condition for the effective action considered as a function of the field $h$ (represents the excitations of the composite scalar field above its condensate).
The value of the Higgs boson mass $M_H^2 \approx m_t^2/2$ follows from the symmetry of the interaction term that takes place at small distances (i.e. large  momenta $|(p-k)^2|, |p^2|, |q^2| \sim M_Z^2 \approx [90 \, {\rm GeV}]^2$). At these values of momenta we have the effective interaction term between the composite Higgs boson and the SM fermions of the form:
\begin{widetext}
\begin{eqnarray}
S_I & = &  - V\sum_{{\bf a}}\sum_{q=p-k}\Bigl[\bar{L}_b^{{\bf a}}(p) R^{\bf a}_U(k)   H^b(q) + \bar{\cal L}_b^{\bf a}(p) {\cal R}^{\bf a}_N(k) H^b(q)  +(h.c.)\Bigr]\label{Sf2222MS} \\&& -V\sum_{{\bf a}}\sum_{q=p-k} \Bigl[\bar{L}_c^{{\bf a}}(k) R^{\bf a}_D(p)  \bar{H}^b(q)\epsilon_{bc} + \bar{\cal L}_c^{\bf a}(k) {\cal R}^{\bf a}_E(p) \bar{H}^b(q)\epsilon_{bc} +(h.c.)\Bigr]  \Bigr], \quad (|(p-k)^2|,|p^2|,|k^2| \sim M_Z^2)\nonumber
\end{eqnarray}
\end{widetext}
All SM fermions enter this interaction term in an equal way. This form of the interaction term points out, that the ultraviolet completion of the Standard Model preserves the symmetry of Eq. (\ref{Sf2222MS}) already at the values of fermion and scalar boson momenta of the order of Electroweak scale $M_Z \sim 90$ GeV. (Recall, that this implies the light fermions to be far out of the mass shell.)
We suppose, that the appearance of the effective interaction of the form of Eq. (\ref{Sf2222MS}) may be checked already using the existing data. At least, we may observe the regime of approaching the effective interaction between the fermions and the Higgs boson to that of Eq. (\ref{Sf2222MS}).

Finally, we would like to remark, that the interaction term of the specific form Eq. (\ref{I4}) suggested here may be too idealized. The actual interaction between the fermions that leads to the formation of their masses may be more complicated. However, we expect that if Nature chooses the pattern of the composite Higgs boson suggested here, the interaction between the composite Higgs boson and the SM fermions should anyway have the form of Eq. (\ref{Sf2222MS}) at small enough distances. This is, probably, the main output of the present paper: we suggest to look carefully at the existing experimental data in order to confirm or reject the existence of the interaction of the form of Eq. (\ref{Sf2222MS}).

\section*{Acknowledgements}

The author is grateful to V.A.Miransky for useful discussions and careful reading of the manuscript, and to G.E.Volovik, who noticed the relation $M_H^2 \approx m_t^2/2$ soon after the discovery of the $125$ GeV Higgs boson. The author is benefited from discussion of the model with Yu. Shylnov, and especially from the discussions with Z.Sullivan of the experimental constraints on the Higgs boson cross sections and the decay branching ratios. The work is supported by the Natural Sciences and Engineering Research Council of
Canada.

\section*{Appendix. Application of zeta - regularization technique to the calculation of various correlators.}

\subsection{Evaluation of the fermion determinant}

Here we use the method for the calculation of various Green functions using zeta - regularization developed in \cite{McKeon}.
In order to calculate the fermion determinant we perform the rotation to Euclidean space - time. It corresponds to the change: $t \rightarrow -i x^4, \bar{\psi} \rightarrow i \bar{\psi}, \gamma^0 \rightarrow \Gamma^4, \gamma^k \rightarrow i \Gamma^k (k = 1,2,3)$. (The new gamma - matrices are Euclidean ones.)
The resulting Euclidean  functional determinant has the form:
\begin{eqnarray}
Z[h] &=& {\rm det} \Bigl[\hat{P}\Gamma + iM + i \hat{G}_h  \Bigr] \\& =& \int D \bar{\psi} D\psi {\rm exp}\Bigl(\int d^4x \bar{\psi}\Bigl[\hat{P}\Gamma + i(M + \hat{G}_h)  \Bigr]\psi \Bigr)\nonumber
\end{eqnarray}
Here $\psi$ involves all SM fermions, $M$ is the fermion mass matrix that originates from the condensate of $h$.
The operator  $\hat{G}_h$ depends on the functions $G(x,y)$ and $h(x)$ and is given by Eq. (\ref{conv}).
 The transformation $\psi \rightarrow \Gamma^5 \psi, \bar{\psi} \rightarrow -\bar{\psi}\Gamma^5$ results in $Z[h] = {\rm det} \Bigl[\hat{P}\Gamma + i(M + \hat{G}_h)  \Bigr] = {\rm det} \Bigl[\hat{P}\Gamma - i(M + \hat{G}_h)  \Bigr]$. Therefore,
\begin{equation}
Z[h] = \Bigl({\rm det} \Bigl[\hat{P}\Gamma + i(M + \hat{G}_h)  \Bigr] \Bigl[\hat{P}\Gamma - i(M + \hat{G}_h)  \Bigr]\Bigr)^{1/2}
\end{equation}
We
get the fermionic part of the Euclidean effective action:
\begin{equation}
S_f[h] = - \frac{1}{2}{\rm Tr} \, {\rm log} \Bigl[\hat{P}^2 + M^2 + (2 M \hat{G}_h +\hat{G}_h^2 -  \Gamma [\partial,\hat{G}_h]) \Bigr]
\end{equation}

We denote $A = \hat{P}^2 + M^2$, and $V = 2 M \hat{G}_h +\hat{G}_h^2 -  \Gamma [\partial, \hat{G}_h]$.
It is worth mentioning, that in momentum space we have
\begin{equation}
[[\partial,\hat{G}_h]\xi](P) =i \sum_{Q = P - K} Q \xi(K)  g(-Q^2,-P^2,-K^2) h(Q) \label{conv2}
\end{equation}

In zeta - regularization we have:
\begin{equation}
S_f[h] = \frac{1}{2} \partial_s \frac{1}{\Gamma(s)} \mu^{2s} \int dt t^{s-1} {\rm Tr} \,{\rm exp}\Bigl[ - (A + V) t\Bigr]
\end{equation}
At the end of the calculation $s$ is to be set to zero.  Here the dimensional parameter $\mu$ appears. We identify this dimensional parameter with the working scale of the interaction that is responsible for the formation of composite $h$ - boson. Further we expand:
\begin{widetext}
\begin{equation}
{\rm Tr} \,{\rm exp}\Bigl[ - (A + V) t\Bigr] = {\rm Tr} \,\Bigl(e^{ - A t} + (-t) e^{ - A t} V + \frac{(-t)^2}{2}\int_0^1 du  e^{ - (1-u)A t}V e^{ - u A t} V + ...\Bigr)\label{expansion}
\end{equation}
\end{widetext}

\subsection{One - point function}
\label{sect1p}
The part of the effective action that produces the one - point Green function of the field $h$ is equal to
\begin{equation}
S_f^{(1)}[h] =  \int {d^4x} {h}(x) C,
\end{equation}
where
\begin{eqnarray}
 C &=&  \partial_s \frac{\Gamma(s+1)}{\Gamma(s)} {\rm Tr}\,\Bigl(- M  \mu^{2s} \int \frac{d^4Q}{(2\pi)^4} \frac{g(0,-Q^2,-Q^2)}{(Q^2 + M^2)^{s+1}} \Bigr)\nonumber\\
 &=& - \frac{2}{8 \pi^2 } \partial_s \sum_{{\bf a}, a} \kappa^{\bf a}_a\frac{[m^{\bf a}_{a}]^{2(1-s)+1}}{s-1}\mu^{2s}
  \nonumber\\&=&   \frac{2}{8 \pi^2 } \sum_{{\bf a}, a} \kappa^{\bf a}_a{[m^{\bf a}_{a}]^3}({\rm log}\frac{\mu^2}{[m^{\bf a}_{a}]^2} + 1)\nonumber\\ &\approx& -\frac{2N_c}{8 \pi^2 } {m_t^3}\Bigl(-{\rm log}\frac{\mu^2}{m_t^2} \Bigr)
\end{eqnarray}
where $\rm Tr$ is over the spinor and flavor indices, $N_c = 3$ is the number of colors. In the first row we take into account, that the momentum circulating within the fermion loop is of the order of the fermion mass $m^{\bf a}_a$. On the language of zeta regularization the interaction scale becomes the parameter entering all expressions to give the dimensionless combinations $m^{\bf a}_a/\mu$. One can easily check that the typical values of $|Q^2|$ are indeed of the order of $[m^{\bf a}_a]^2$ as follows.
We use the standard expression for the loop integral:
\begin{eqnarray}
J &=& \int  \frac{d^4Q}{(2\pi)^4} \frac{\mu^{2s}}{(Q^2 + M^2)^{s+1}} \nonumber\\ &=&  \frac{1}{16 \pi^2} M^2 \Bigl(\frac{\mu^2}{M^2}\Bigr)^s \frac{\Gamma(s-1)}{\Gamma(s+1)}
\end{eqnarray}
The typical value of momentum $|Q| = \sqrt{Q^2}$ circulating within the loop may be estimated as  $J \sim \frac{Q^{4}\mu^{2s}}{(Q^2+M^2)^{s+1}} \sim \mu^{2s}Q^{2-2s} \sim M^2 \Bigl(\frac{\mu^2}{M^2}\Bigr)^s$. From here we derive $Q \sim M$.  This distinguishes dramatically zeta regularization from the usual cutoff regularization, where the typical value of momentum circulating within the divergent fermion loop becomes of the order of the ultraviolet cutoff instead of the natural internal parameter of the theory (such as the fermion mass).

As a result we substitute $g(0,-Q^2,-Q^2) \approx \kappa$ according to our basic supposition about the Formfactors $g$.

\subsection{Two - point Green function in Euclidean region}
\label{secttwopoint}
The part of the effective action that produces the two - point Green functions of the field $h$ is equal to
\begin{equation}
S_f^{(2)}[h] = V \int \frac{d^4 P}{(2\pi)^4} h(P) \Pi(-P^2) h(-P),
\end{equation}
where $\Pi = \Pi^{(1)} + \Pi^{(2)} + \Pi^{(3)}$. Below we calculate these three terms separately. In this subsection we shall be interested in the values of $\Pi(-P^2)$ in the Euclidean region $P^2 \sim M_H^2$ that correspond to the space - like four - momentum of the Higgs field in space - time of Minkowsky signature. This will be required for the estimate of the values of the form - factors $g^{\bf a}_a$ in what follows.
 In the next subsection we shall discuss the analytical continuation of the obtained results to the negative values of $P^2$ that correspond to the time - like four - momenta of the field $h$.

$\Pi^{(1)}$ corresponds to the term in $V$ proportional to $\hat{G}_h$ squared.
\begin{widetext}
\begin{eqnarray}
\Pi^{(1)}(-P^2) &=& \frac{1}{2} \partial_s \frac{\Gamma(s+1)}{\Gamma(s)} {\rm Tr}\,\Bigl(- \int  \frac{d^4Q}{(2\pi)^4} \frac{|g(-P^2, -Q^2, - (P+Q)^2)|^2}{(Q^2 + M^2)^{s+1}}\mu^{2s} \Bigr)\nonumber\\
 &\approx & - \frac{1}{8 \pi^2 } \partial_s \frac{[m_t]^{2(1-s)}}{s-1} \approx   \frac{1}{8 \pi^2 } {[m_{t}]^{2}}({\rm log}\frac{\mu^2}{[m_{t}]^2} + 1)\label{P1}
\end{eqnarray}
\end{widetext}
where $\rm Tr$ is over the spinor and flavor indices. Here the top quark dominates the sum, which follows from the properties of the Form - factors $g$ because within the integral for the light fermions the typical values of momenta dominate that are much smaller, than $m_t$. To see this we recall that $|g(-P^2,-(P+Q)^2,-Q^2)| \le 1$ for each fermion. Therefore the absolute value of the contribution of each fermion with mass $m_f$ is not larger, than  $\sim \frac{1}{8 \pi^2 } {m_f^{2}}({\rm log}\frac{\mu^2}{m_f^2} + 1)$.

$\Pi^{(2)}$ corresponds to $V^2$ and to the term in $V$ proportional to $-\Gamma[\partial, \hat{G}_h]$.
\begin{widetext}
\begin{eqnarray}
\Pi^{(2)}(-P^2) &=&  \partial_s \frac{1}{4\Gamma(s)} \int dt du t^{s+1}{\rm Tr}\,\Bigl( \Gamma^{\mu} \Gamma^{\nu} P_{\mu}P_{\nu}\int \frac{d^4Q}{(2\pi)^4} |g^{\bf a}_{a}(-P^2, -Q^2,-(P+Q)^2)|^2\nonumber\\&&  e^{- t (Q^2 u + (P+Q)^2(1-u) + M^2)}  \mu^{2s}\Bigr)\nonumber\\
&=&  \partial_s \frac{1}{\Gamma(s)} \int dt du t^{s+1}\sum_{{\bf a}, a} \,\Bigl(P^2\int \frac{d^4Q}{(2\pi)^4} |g^{\bf a}_{a}(-P^2,-(Q + u P - P)^2,-(Q +u P)^2)|^2 \nonumber\\&&e^{- t (P^2 u(1-u) + Q^2 + [m^{\bf a}_{a}]^2)}  \mu^{2s}\Bigr)\nonumber\\
&\approx &  \partial_s \frac{ \Gamma(s+2)}{\Gamma(s)} \int du \sum_{{\bf a},a} \,\Bigl( P^2\int \frac{d^4Q}{(2\pi)^4} \frac{|g^{\bf a}_{a}(-P^2,-(Q + u P - P)^2,-(Q +u P)^2)|^2}{(P^2 u(1-u) + Q^2 + [m^{\bf a}_{a}]^2)^{s+2}} \mu^{2s} \Bigr) \label{P20}
\end{eqnarray}
\end{widetext}
We applied the transformation $Q \rightarrow Q - (1-u)P$. In the fermion loop in $\Pi^{(2)}$ the typical values of circulating momentum $Q^2$ are of the order of $P^2$. This may be demonstrated in the similar way as done in Sect. \ref{sect1p}. Namely, in the region $Q^2 \sim P^2 \sim M_H^2$ also the typical values of $(Q + u P - P)^2$ and $(Q + u P)^2$ are of the order of $\sim M_H^2$. Therefore, we substitute $g$ by unity into the integral for light fermions, and have the integral of the form
\begin{eqnarray}
&&J(-P^2)  =  \int  \frac{d^4Q}{(2\pi)^4} \frac{\mu^{2s}}{(Q^2 + P^2 u(1-u) +  [m^{\bf a}_{a}]^2)^{s+2}} \nonumber\\ &&\approx \frac{1}{16 \pi^2} M^2 {\mu^{2s}}{\Bigr(P^2 u(1-u) +  [m^{\bf a}_{a}]^2\Bigr)^{-2s}} \frac{\Gamma(s)}{\Gamma(s+2)}\nonumber\\ &&\approx   \frac{1}{16 \pi^2} M^2 {\mu^{2s}}{\Bigr(P^2 u(1-u) \Bigr)^{-2s}} \frac{\Gamma(s)}{\Gamma(s+2)}
\end{eqnarray}
Here we omit the fermion masses because for the light fermions $m^2$ is much smaller, than $P^2 \sim M_H^2$. 
On the other hand $J$ may be expressed through the typical value of momentum $Q$ as $J \sim \frac{\mu^{2s}}{Q^{2s}}$. From here we derive the typical value of momentum $Q^2 \sim P^2 u(1-u) \sim P^2$.

This justifies the supposition, that all form - factors $G$ may be substituted by unity in the expression for $\Pi^{(2)}$.
The other regions of momenta $Q$, where the values of the form - factors $g$ are much smaller than $1$ do not contribute essentially to the overall integral.
 Thus, we substitute $g^{\bf a}_{a}(-P^2, -(-Q - u P +P)^2,-(Q+u P)^2) \approx g(-M_H^2,-M_H^2,-M_H^2) \approx 1$ for all fermions. This gives the estimate
$\Pi^{(2)}(-P^2)
 \approx    \frac{1}{16 \pi^2}  \sum_{{\bf a},a} \,\Bigl(P^2 \int_0^1 du \,{\rm log}\, \frac{\mu^2}{P^2 u(1-u) + [\tilde{m}^{\bf a}_a]^2}  \Bigr) $. 
Here $\tilde{m}^{\bf 3}_U = m_t$ while the other values of mass parameters differ from the masses of the light fermions. Instead we should substitute the values of mass parameters such that for $|Q| \gg \tilde{m}^{\bf a}_a$ we have $g^{\bf a}_a(M_H^2,Q^2,Q^2) \approx 1$ while for $|Q| \ll \tilde{m}^{\bf a}_a$ we have $g^{\bf a}_a(M_H^2,Q^2,Q^2) \ll 1$ (the final answer will not depend on the particular values of these mass parameters).  Finally, we arrive at
\begin{eqnarray}
\Pi^{(2)}(-P^2)
 &\approx &    \frac{N_{\rm total}}{16 \pi^2}  \, P^2  \,{\rm log}\, \frac{\mu^2}{m_t^2} \label{P2}
\end{eqnarray}

  The last term corresponds to $V^2$ and to the contribution of $2 M \hat{G}_h$ in $V$. It is evaluated for any values of $P^2$ in the way similar to that of $\Pi^{(1)}$:
\begin{widetext}
\begin{eqnarray}
\Pi^{(3)}(-P^2) &=&  \partial_s \frac{1}{\Gamma(s)} \int dt du t^{s+1}{\rm Tr}\,\Bigl( M^2 \int \frac{d^4Q}{(2\pi)^4} |g^{\bf a}_{a}(-P^2, - Q^2, -(P+Q)^2)|^2 e^{- t (Q^2 u + (P+Q)^2(1-u) + M^2)}  \Bigr)\nonumber\\
 &\approx&   4 N_c m_t^2  \frac{1}{16 \pi^2}  \, \int_0^1 du \,{\rm log}\, \frac{\mu^2}{P^2 u(1-u) + m_t^2} \approx 4 N_c m_t^2 \frac{1}{16 \pi^2}  \, {\rm log}\, \frac{\mu^2}{m_t^2}\label{P3}
\end{eqnarray}
\end{widetext}
for $\mu \gg m_t$.

Total expression for the function $\Pi(-P^2)$ for $P^2 \sim M_H^2$ and $\mu \gg m_t$ has the form:
\begin{widetext}
\begin{eqnarray}
\Pi(-P^2)  \approx   (N_{\rm total}P^2 + 4 N_c m_t^2) \, \frac{1}{16 \pi^2}  \, {\rm log}\, \frac{\mu^2}{m_t^2}  + \frac{1}{8 \pi^2} N_c{m_t^{2}}({\rm log}\frac{\mu^2}{m_t^2} + 1)
\end{eqnarray}
\end{widetext}
Here $m_t \approx 174$ GeV is the top - quark mass,  $N_{\rm total} = 2(3N_c + 3)  = 24 $ is the total number of SM quarks and leptons, while $N_c = 3$ is the number of colors.

\subsection{Two - point Green function in space - time of Minkowsky signature}

Let us consider the two - point correlation function for the Higgs boson in Minkowsky space - time $\Pi(p^2) = \Pi^{\prime}(p^2) + i \Pi^{\prime\prime}(p^2)$ that depends on the external momentum of the Higgs boson $p^2$. It is related to the action in Minkowsky space - time as $S = - \int d^4x h(x)\Pi(\hat{p}^2) h(x)$. This quantity may be represented as the Euclidean loop integral for the external Euclidean momentum $P^2$. For the simplicity we set ${\bf P} = 0$ and $P^2 = P_4^2$.

The consideration of the parts $\Pi^{(1)}, \Pi^{(3)}$ is the most simple one. The top - quark dominates in the resulting expressions for all values of $P^2$. The result does not depend on momentum and is given by Eq. (\ref{P1}) and Eq. (\ref{P3}).

In consideration of $\Pi^{(2)}$  we notice, that for $P^2 \sim M_H^2$ the fermion Euclidean momenta $Q,P+Q$ with $(Q+P)^2 \sim Q^2 \sim P^2$ dominate. Then we substitute into the loop integral   $g^{\bf a}_{a}(-P^2,-Q^2,-(P+Q)^2) = 1$. The resulting real - valued expression $\Pi^{(2)}(-P^2)$ depends on the external Euclidean momentum $P^2>0$. However, we shall be interested in the complex - valued expression for negative values  $P^2 = -p^2 \sim -M_H^2$. In order to construct the analytical continuation of the expression $\Pi^{(2)}(-P^2)$ with $P^2 \sim + M_H^2$ to $P^2 \sim -M_H^2$ we should choose the line in the complex plane of $P_4$ that connects points $M_H$ and $-i M_H$, and along which the analytical continuation is to be performed. We choose the sector of the circle with $|P| = M_H$. While moving along this line for $P \sim e^{i \phi} M_H$ we rotate in the integral $Q_4 \rightarrow e^{i \phi} |Q_4|$. It occurs, that after this rotation the values of the momenta $|Q| \sim M_H$ (and as a consequence $|P+Q|\sim M_H$) remain dominant in the real part $\Pi^{(2)\prime}$ of $\Pi^{(2)}$, then we may simply substitute into the approximate analytical Euclidean expression $\Pi^{(2)}(-P^2)$ (calculated with $g^{\bf a}_{a}(-P^2,-Q^2,-(P+Q)^2) = 1$) the value $P^2 \sim - M_H^2$, take its real part and obtain the needed expression for $\Pi^{(2)\prime}(p^2)$ with $p^2 \sim M_H^2$ in space of Minkowsky signature.
In the limit, when the new strong interaction scale given by $\mu$ is much larger, than all fermion masses we combine the three terms $\Pi^{(1)},\Pi^{(2)},\Pi^{(3)}$ and arrive at:
\begin{eqnarray}
\Pi^{\prime}(p) &\approx &  -   (p^2 - \frac{4 m_t^2 N_c}{N_{\rm total}}) \,{Z_h^2}  + \frac{N_c}{8 \pi^2}   \, m_t^2  \, {\rm log}\frac{\mu^2}{m_t^2}\nonumber\\&&  \quad {\rm for}\, |p^2| \sim M_H^2\label{wyA}
\end{eqnarray}
with
\begin{equation}
{Z_h^2} \approx \frac{N_{\rm total}}{16 \pi^2} \, {\rm log}\frac{\mu^2}{m_t^2} \label{Zhl}
\end{equation}
We may check ourselves and estimate expression for $\Pi(p^2)$ for the form - factors of the form of Eq. (\ref{ex2}). The dominant contribution to $\Pi^{\prime}$ is given by Eq. (\ref{wyA}) while the sub - dominant terms are given by the residues at the positions of the poles of Eq. (\ref{ex2}). The resulting sub - dominant contributions are suppressed by the ratios $\sim \sqrt{\lambda^{\bf a}_a} M_I^2/M_H^2 \ll 1$.

For the imaginary part $\Pi^{\prime \prime}$ of $\Pi$ the situation is different: say, for $iP_4 = M_H - i 0$ in the integral over $Q$ for $\Pi^{\prime \prime}$ the values  $Q^2 = -m_f^2$ and $(Q+P)^2 = -m_f^2$ dominate (this corresponds to the decay of the Higgs boson into the pair fermion - antifermion with masses $m_f$ and $4$ - momenta $q^2 = (p+q)^2 = +m_f^2$).  As a result in order to calculate the imaginary part of $\Pi(p^2)$ we cannot simply substitute $P^2 \sim -M_H^2$ into the Euclidean expression. The consideration is more involved and we are to substitute   $g^{\bf a}_{a}(M_H^2,m_f^2,m_f^2)$ instead of $g^{\bf a}_{a}(-P^2,-Q^2,-(P+Q)^2) = 1$. This gives
\begin{widetext}
\begin{eqnarray}
\Pi^{\prime\prime}(p^2)  \approx  \frac{1}{16 \pi} \sum_{{\bf a},a}(p^2 - 4 [m^{\bf a}_a]^2)|g^{\bf a}_{a}(M_H^2,[m^{\bf a}_a]^2,[m^{\bf a}_a]^2)|^2\theta(p^2 - 4 [m^{\bf a}_a]^2)
\end{eqnarray}
\end{widetext}
Since all values $\kappa^{\bf a}_a \ll 1$ for all fermions except for the top quark, we may neglect the imaginary part of $\Pi$ at $p^2 < 4 m_t^2$.

Notice, that the result for the one - point function obtained above as well as the expressions for $\Pi^{(1)}, \Pi^{(3)}$ are valid for all possible values of the momentum of the field $h$. At the same time the term proportional to $p^2$ in  Eq. (\ref{wyA}) (that originates from $\Pi^{(3)}$) is valid for the values of momenta of the order of the Higgs boson mass $|p^2| \sim M_H^2$. In order to obtain the expression for the effective action of the Higgs boson field $h$ for all values of momenta $p$ we need to modify this expression. The local effective action for the field $h$ will result in the following expression for $\Pi(p)$ valid for all values of momentum $p$:
\begin{equation}
\Pi(p) \approx   -Z^2_h(w(p^2) p^2 - M_{H}^2)  + \frac{N_c}{8 \pi^2}   \, m_t^2  \, {\rm log}\frac{\mu^2}{m_t^2}, \label{wyA1}
\end{equation}
where $Z^2_h = \frac{N_{\rm total}}{16 \pi^2}{\rm log}\, \frac{\mu^2}{m_t^2}$, while $w(p^2)\approx 1$ for $|p^2| \sim M_H^2$, and
\begin{eqnarray}
M_{H}^2 \approx 4 N_c m_t^2/N_{\rm total}
\end{eqnarray}

We may easily estimate the values of $w(p^2)$ for $|p^2| \ll M_H^2$. The expressions for $\Pi^{(1)}$ and $\Pi^{(3)}$ remain the same while the expression for $\Pi^{(2)}$ is given by the integral considered in the previous subsection with the typical values $Q^2 \ll M_H^2$ instead of $Q^2 \sim M_H^2$. That's why we substitute
$g^{\bf a}_{a}(0,0,0) \approx \kappa^{\bf a}_a$ instead of $g^{\bf a}_{a}(-P^2,-Q^2,-(P+Q)^2) = 1$. This leads to
\begin{eqnarray}
&& w(p^2) \approx \frac{N_c}{N_{\rm total}} = 1/8 \quad {\rm for}\, |p^2| \ll M_Z^2\nonumber
\end{eqnarray}

The expression of $w(p^2)$ for the intermediate values of momenta depends on the particular form of the form - factors $g$. However, in order to evaluate the main physical quantities (masses of Higgs boson, Z boson and W boson) we do not need to know the particular form of this function for all ranges of momenta.

\end{document}